\documentclass[a4paper,11pt]{article}
\usepackage{pos}

\title{Calculation of kaon semileptonic form factor with the PACS10 configuration}

\author*[a,b]{Takeshi Yamazaki}

\author[c]{Ken-ichi~Ishikawa}

\author[b]{Naruhito~Ishizuka}

\author[b]{Yoshinobu~Kuramashi}

\author[d]{Yoshifumi~Nakamura}

\author[e]{Yusuke~Namekawa}

\author[b]{Yusuke~Taniguchi}

\author[b]{Naoya~Ukita}

\author[b]{Tomoteru~Yoshi\'e}

\affiliation[]{\normalsize{\bf \sffamily \hspace{50mm} (PACS Collaboration)}}

\affiliation[a]{Faculty of Pure and Applied Sciences, University of Tsukuba, Tsukuba, Ibaraki 305-8571, Japan}

\affiliation[b]{Center for Computational Sciences, University of Tsukuba, Tsukuba, Ibaraki 305-8577, Japan}

\affiliation[c]{Core of Research for the Energetic Universe, Graduate School of Advanced Science and Engineering, Hiroshima University, Higashi-Hiroshima, 739-8526, Japan}

\affiliation[d]{RIKEN Center for Computational Science, Kobe, Hyogo 650-0047, Japan}

\affiliation[e]{Yukawa Institute for Theoretical Physics, Kyoto University, Kyoto 606-8502, Japan}

\emailAdd{yamazaki@het.ph.tsukuba.ac.jp}

\abstract{
We present preliminary results for the kaon semileptonic form factors
using the PACS10 configurations, whose physical volume is more than
(10 fm)$^3$ at the physical point with the lattice spacings of 
0.085 and 0.064 fm.
The configurations were generated using the Iwasaki gauge action and $N_f=2+1$ 
stout-smeared Clover quark action. 
For the continuum extrapolation, we calculate the form factors 
with the local and conserved vector currents.
The form factors in the two lattice spacings are extrapolated 
to the continuum limit using a fit function
based on the NLO SU(3) ChPT formula with terms corresponding to finite lattice
spacing effects.
The value of $|V_{us}|$ is determined
using our preliminary result of the form factor at the zero momentum 
transfer in the continuum limit. 
The result is compared with recent lattice results,
and also predictions of the standard 
model from the unitarity of the Cabibbo-Kobayashi-Maskawa (CKM) matrix.
}

\FullConference{%
 The 38th International Symposium on Lattice Field Theory, LATTICE2021
  26th-30th July, 2021
  Zoom/Gather@Massachusetts Institute of Technology
}

\begin{document}
\maketitle

\section{Introduction}

One of urgent tasks for the particle physics is to search for
signals of physics beyond the standard model.
$|V_{us}|$, which is one of the CKM
matrix elements, is a candidate of the signal,
because recently it was reported~\cite{Bazavov:2018kjg} 
that there is a deviation of $|V_{us}|$ between 
the experimental values determined from the kaon decays, 
especially the semileptonic ($K_{l3}$) decay, and 
an estimation with the unitarity condition of the CKM matrix.

For a precise determination of $|V_{us}|$ through 
the $K_{l3}$ decay process,
it is important to calculate the $K_{l3}$ form factor 
at the zero momentum transfer squared, $q^2 = 0$, in lattice QCD calculation,
because other quantities for $|V_{us}|$ are already determined in the experiment accurately~\cite{Moulson:2017ive}.
So far several lattice calculations for the $K_{l3}$ form factor~\cite{Dawson:2006qc,Boyle:2007qe,Lubicz:2009ht,Bazavov:2012cd,Boyle:2013gsa,Bazavov:2013maa,Boyle:2015hfa,Carrasco:2016kpy,Aoki:2017spo,Bazavov:2018kjg}
were carried out.
In this work, for a precise measurement of the form factor and
to confirm the deviation of $|V_{us}|$,
we perform the calculation using the PACS10 configurations,
whose spatial extent is more than 10 fm at the physical point.
Using the configuration, finite volume
effects are considered to be negligible, and the chiral extrapolation
of the form factor is not necessary.
Thus, the configuration is suitable for 
precise measurements of the physical quantities.
The results with the PACS10 configuration at the lattice spacing $a = 0.085$ fm
were already reported in Ref.~\cite{PACS:2019hxd}.
In this report we will show updated results with another PACS10 configuration 
of $a = 0.064$ fm.
All the results presented in this report are preliminary.

\section{Simulation parameters}

For the generation of the PACS10 configurations,
we employ the Iwasaki gauge action~\cite{Iwasaki:2011jk}
and the six stout-smeared Clover quark action in the $N_f=2+1$ QCD. 
The masses for the light and strange quarks are tuned to be the physical ones.
In addition to the ensemble at $a = 0.085$ fm~\cite{Ishikawa:2019qwn} used 
in our previous calculation~\cite{PACS:2019hxd},
another ensemble at $a = 0.064$ fm is used in this work.
The parameters for both ensembles are tabulated
in Table~\ref{tab:sim_param}.

The calculation method of the $K_{l3}$ form factors is basically the same 
in both the lattice spacings, where we evaluate 
the pion and kaon two-point functions and the $K_{l3}$ 
three-point function, and extract appropriate matrix elements
for the form factors.
Details of the method are explained in Ref.~\cite{Ishikawa:2019qwn} 
for the $a = 0.085$ fm case.
At $a = 0.064$ fm, 
we adopt the hopping parameters for the light and strange quarks,
$(\kappa_l, \kappa_s) = (0.125814, 0.124925)$,
and the improved Clover coefficient $c_{\rm SW} = 1.02$,
which is non-perturbatively determined.
We calculate the correlation functions with 
the $Z(2)\otimes Z(2)$ random source~\cite{Boyle:2008yd}.
An exponential smeared source with the random source 
is also employed at $a = 0.064$ fm
to investigate the source dependence of the result.
We calculate the three-point functions
with several temporal source and sink separations
from 3.1 to 4.1 fm for the random source, and from 2.3 to 3.5 fm for
the exponential smeared source, respectively.
In order to take the continuum limit,
the point splitting conserved vector current is employed
for the $K_{l3}$ three-point function as well as the local vector current
in the two lattice spacings.
The statistical error in the calculation is evaluated by the jackknife method.

\begin{table}[!h]
\caption{
Simulation parameters of the PACS10 configurations in the two lattice spacings.
The bare coupling ($\beta$), lattice size ($L^3\cdot T$),
physical spatial extent ($L$[fm]), pion and kaon masses ($M_\pi$, $M_K$)
are tabulated.
$N_{\rm conf}$ and $N_{\rm meas}$ represent the number of the configurations
and the maximum number of the measurement, respectively.
  \label{tab:sim_param}
}
\begin{center}
\begin{tabular}{cccccccc}\hline\hline
$\beta$ & $L^3\cdot T$ & $L$[fm] & $a$[fm] & $M_\pi$[MeV] & $M_K$[MeV] &
$N_{\rm conf}$ & $N_{\rm meas}$
\\\hline
1.82 & 128$^4$ & 10.9 & 0.085 & 135 & 497 & 20 & 1280 \\
2.00 & 160$^4$ & 10.2 & 0.064 & 137 & 501 & 20 & 1280 \\\hline\hline
\end{tabular}
\end{center}
\end{table}

\section{Results}

\subsection{$K_{l3}$ form factors at $a = 0.064$ fm}

In this subsection we present preliminary results of 
the $K_{l3}$ form factors at $a = 0.064$ fm.

The matrix element for the $K_{l3}$ form factors given by
\begin{equation}
\langle \pi(p)|V_\mu | K(0)\rangle = (p_K+p_\pi)_\mu f_+(q^2)
+ (p_K - p_\pi)_\mu f_-(q^2)
\label{eq:mat_ele_kl3} ,
\end{equation}
where $q^2 = -(M_K-E_\pi)^2 + \vec{p}^2$,
is extracted from the three-point functions using a proper fit form
including excited state effects for $\pi$ and $K$
as described in Ref.~\cite{PACS:2019hxd}.
The temporal wrapping around effects for the mesons
are suppressed by averaging the three-point functions with
the periodic and anti-periodic boundary conditions 
in the temporal direction~\cite{PACS:2019hxd}.
Since the results from the random and exponential smeared sources are 
in good agreement with each other, we carry out a combined fit analysis
using both the data to extract the matrix elements.

Using the extracted matrix elements, 
the two form factors, $f_+(q^2)$ and $f_0(q^2)$,
are obtained by solving linear equations of Eq.~(\ref{eq:mat_ele_kl3})
at each $q^2$.
The definition of $f_0(q^2)$ is given by
\begin{equation}
f_0(q^2) = f_+(q^2) - \frac{q^2}{M_K^2 - M_\pi^2}f_-(q^2) .
\end{equation}
The two form factors coincide at $q^2 = 0$, $f_+(0) = f_0(0)$,
by the definition.

\begin{figure}[!th]
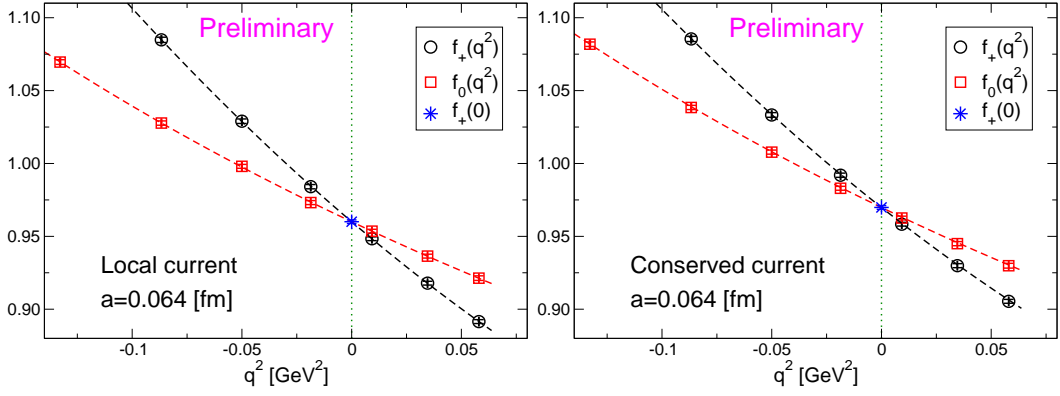

 \centering
 \includegraphics*[scale=0.40]{figs/fig-fit-kl3_lcl.eps}
 \includegraphics*[scale=0.40]{figs/fig-fit-kl3_cns.eps}
 \caption{
The $K_{l3}$ form factors at $a = 0.064$ fm with the local (left)
and conserved (right) currents as a function of $q^2$. 
The circle and square symbols represent
the data for $f_+(q^2)$ and $f_0(q^2)$, respectively.
The dashed curves express their interpolations to $q^2 = 0$ using
the fit functions based on 
the NLO SU(3) ChPT formulas in Eqs.~(\ref{eq:NLO_chpt_f+}) 
and (\ref{eq:NLO_chpt_f0}).
The fit result of $f_+(0)$ is also plotted by the star symbol.
  \label{fig:ffs_b200}
 }
\end{figure}

The preliminary results for $f_+(q^2)$ and $f_0(q^2)$
with the local and conserved currents
are presented in Fig.~\ref{fig:ffs_b200} as a function of $q^2$.
Thanks to the large spatial extent in our calculation,
we obtain the form factors in a tiny $q^2$ region with the periodic
boundary condition in the spatial directions.
The value of $f_+(0)$, which is an essential quantity to determine
$|V_{us}|$, is obtained by an interpolation of the form factors to $q^2 = 0$.

We perform the interpolation for each current data independently
in Fig.~\ref{fig:ffs_b200}.
The dashed curves in the figure represent simultaneous fit results 
using the data for $f_+(q^2)$ and $f_0(q^2)$.
As in our previous work~\cite{PACS:2019hxd},
the NLO SU(3) ChPT formulas~\cite{Gasser:1984gg,Gasser:1984ux} 
are employed in each form factor given by
\begin{eqnarray}
f_+(q^2) &=& 1 - \frac{4}{F_0^2} L_9 q^2 + K_+(q^2) + c_0 + c_2^+ q^4 ,
\label{eq:NLO_chpt_f+}\\
f_0(q^2) &=& 1 - \frac{8}{F_0^2} L_5 q^2 + K_0(q^2) + c_0 + c_2^0 q^4 ,
\label{eq:NLO_chpt_f0}
\end{eqnarray}
where $L_9$, $L_5$, $c_0$, and $c^{+,0}_{2}$ are free parameters.
The last two terms in the equations correspond to a part of 
the NNLO analytic terms.
A common $c_0$ is adopted in the two fit forms to incorporate
the constraint at $q^2 = 0$, $f_+(0) = f_0(0)$.
The pion decay constant in the chiral limit is fixed to $F_0 = 0.11205$ GeV
as in our previous calculation.
The two functions $K_+(q^2)$ and $K_0(q^2)$ 
depend on $M_\pi$ and $M_K$, whose explicit forms are
shown in Ref.~\cite{PACS:2019hxd}.
The simultaneous fits work well in both the current data.
The result of $f_+(0)$ is expressed by the star symbol in the figure.
While it is hard to see in the figure, we observe a clear difference 
of the value of $f_+(0)$ with the local and conserved currents.
This difference stems from a finite lattice spacing effect,
and can be used for the continuum extrapolation of $f_+(0)$.

\subsection{Continuum extrapolation of form factors}

\begin{figure}[!th]
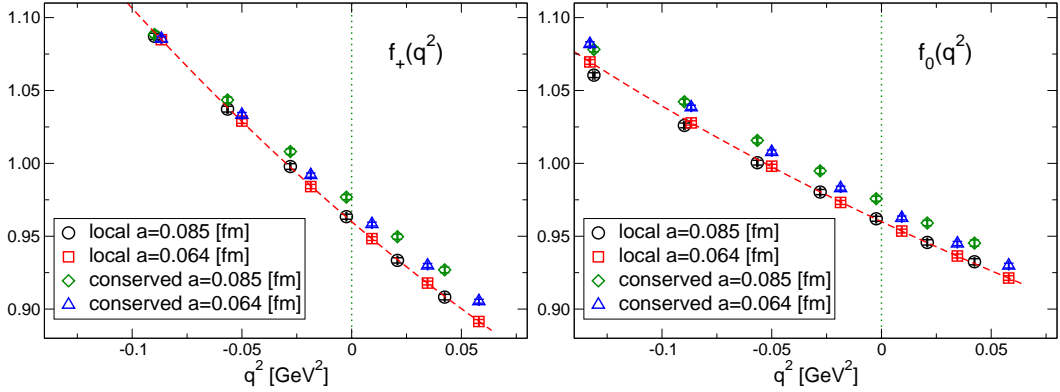

 \centering
 \includegraphics*[scale=0.40]{figs/fplus_q2.eps}
 \includegraphics*[scale=0.40]{figs/fzero_q2.eps}
 \caption{
All the data for the $K_{l3}$ form factors, $f_+(q^2)$ (left)
and $f_0(q^2)$ (right), in our calculation as a function of $q^2$.
The circle and square denote
the local current data at $a = 0.085$ and 0.064 fm, respectively.
The diamond and triangle express those for the conserved current data.
The dashed curve, which is the fit results of the local current
data at $a = 0.064$ fm, is also plotted for comparison.
  \label{fig:ffs_summary}
 }
\end{figure}

Figure~\ref{fig:ffs_summary} presents all the data
in our calculation for $f_+(q^2)$ and $f_0(q^2)$
in the two different lattice spacings with the local and 
conserved currents.
The dashed curve is the fit result at $a = 0.064$ fm
using the local current data explained in the last subsection.

The local current data of $f_+(q^2)$ at $a = 0.085$ fm is well consistent with
the dashed curve in all the $q^2$ region.
It suggests that a finite lattice spacing effect is small
in $f_+(q^2)$ with the local current.
While the difference between the local and conserved current data 
increases with $q^2$ at each lattice spacing, 
the conserved current data approach to the dashed curve
as the lattice spacing decreases.

The lattice spacing dependence of $f_0(q^2)$ is more complicated 
than the one in $f_+(q^2)$.
Nevertheless, the discrepancy between the local and conserved current data 
decreases with the lattice spacing in all the $q^2$ region.
In the large $q^2$ region, all the data converge the dashed curve
obtained from the local current data.
In contrast to the large $q^2$,
the data in the small $q^2$ region seem to approach the conserved current data 
as the lattice spacing decreases.

\begin{figure}[!th]
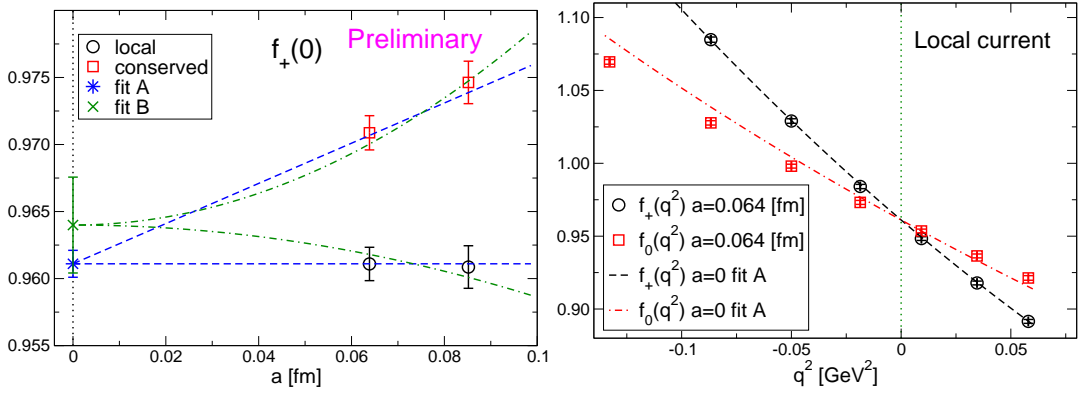

 \centering
 \includegraphics*[scale=0.40]{figs/f00_a0.eps}
 \includegraphics*[scale=0.40]{figs/fplus_fzero_a0.eps}
 \caption{
Left: The lattice spacing dependence of $f_+(0)$ with the local (circle)
and conserved (square) current data. The continuum extrapolations
and the results in the continuum limit with the fit A and B in
Eqs.~(\ref{eq:ato0_NLO_chpt_f+})--(\ref{eq:ato0_NLO_chpt_f0_fitB})
are also plotted.
Right: The continuum results of the fit A 
for $f_+(q^2)$ and $f_0(q^2)$
are represented by the dashed and dot-dashed curves, respectively,
as a function of $q^2$.
The local current data at $a = 0.064$ fm are also plotted for comparison.
  \label{fig:ato0_f00_fplus_fzero}
 }
\end{figure}

The left panel of Fig.~\ref{fig:ato0_f00_fplus_fzero} shows
the results of $f_+(0)$ obtained from the fits based on 
the NLO SU(3) ChPT formulas
in Eqs.~(\ref{eq:NLO_chpt_f+}) and (\ref{eq:NLO_chpt_f0})
in each lattice spacing
with the local and conserved currents as a function of the lattice spacing.
While the local current results are reasonably flat against the lattice
spacing, the conserved ones show a clear dependence on the lattice spacing.

We carry out simultaneous continuum extrapolations with all the data
we obtained.
The fitting forms are based on the NLO SU(3) ChPT formulas in
Eqs.~(\ref{eq:NLO_chpt_f+}) and (\ref{eq:NLO_chpt_f0}),
and we add terms corresponding to finite lattice spacing effects,
which depend on the local and conserved currents as given by
\begin{eqnarray}
f_+^{\rm cur}(q^2) &=& 
1 - \frac{4}{F_0^2} L_9 q^2 + K_+(q^2) + c_0 + c_2^+ q^4
+ g^{\rm cur}_+(q^2,a),
\label{eq:ato0_NLO_chpt_f+}\\
f_0^{\rm cur}(q^2) &=& 
1 - \frac{8}{F_0^2} L_5 q^2 + K_0(q^2) + c_0 + c_2^0 q^4
+ g^{\rm cur}_0(q^2,a),
\label{eq:ato0_NLO_chpt_f0}
\end{eqnarray}
where ${\rm cur} = {\rm loc, con}$ correspond to the local and 
conserved currents, respectively.

From the above observations of the lattice spacing dependence
for the form factors,
we empirically choose the functions $g_{+,0}^{\rm cur}(q^2,a)$ given as
\begin{eqnarray}
&&g_+^{\rm loc}(q^2) = 0, \ \ g_0^{\rm loc}(q^2) = d_1^0 a^2 q^2, 
\label{eq:ato0_NLO_chpt_f+_fitB}\\
&&g_+^{\rm con}(q^2) = e_0 a + e_1^+ a q^2, \ \ 
g_0^{\rm con}(q^2) = e_0 a + e_1^0 a q^2 ,
\label{eq:ato0_NLO_chpt_f0_fitB}
\end{eqnarray}
where $d_1^0$, $e_0$, $e_1^{+,0}$ are free parameters.
We shall call this fit as ``fit A'' in the following.
In this choice, we assume that the conserved current data have $O(a)$ errors,
which is expected from the lattice spacing dependence of $f_+(0)$ 
shown in the left panel of Fig.~\ref{fig:ato0_f00_fplus_fzero}.
From the fit we obtain the result of $f_+(0)$ denoted by the star symbol
in the panel,
whose fit lines are expressed by the dashed lines.
The fit results for the form factors in the continuum limit 
as a function of $q^2$ are plotted in the right panel in the figure
together with the local current data at $a = 0.064$ fm for comparison.
As discussed in Fig.~\ref{fig:ffs_summary}, the fit result of $f_0(q^2)$ 
is slightly different from the data at $a = 0.064$ fm.
In the continuum limit results,
we use the experimental values for $M_\pi$ and $M_K$ 
instead of their measured values.

In order to estimate systematic error coming from a choice of the fitting
form of the continuum extrapolation, we carry out another fit,
where we choose the functions $g_{+,0}^{\rm cur}(q^2,a)$ as
\begin{eqnarray}
&&g_+^{\rm loc}(q^2) = d_0 a^2, \ \ g_0^{\rm loc}(q^2) = d_0 a^2 + d_1^0 a^2 q^2, \\
&&g_+^{\rm con}(q^2) = e_0 a^2 + e_1^+ a^2 q^2, \ \ 
g_0^{\rm con}(q^2) = e_0 a^2 + e_1^0 a^2 q^2 ,
\end{eqnarray}
with free parameters, $d_0$, $d_1^0$, $e_0$, and $e_1^{+,0}$.
This choice is called as ``fit B''.
The reason of this choice is that our improved coefficient $c_{\rm SW}$
is non-perturbatively determined, so that it is simply considered that
finite lattice spacing effects start from $O(a^2)$.
The fit result of $f_+(0)$ is drawn by
the dot-dashed curves in the left panel of Fig.~\ref{fig:ato0_f00_fplus_fzero}.
This fit result in the continuum limit denoted by the cross symbol agrees with 
the one from the fit A, albeit it has larger error.

\subsection{Result in the continuum limit}

\begin{figure}[!th]
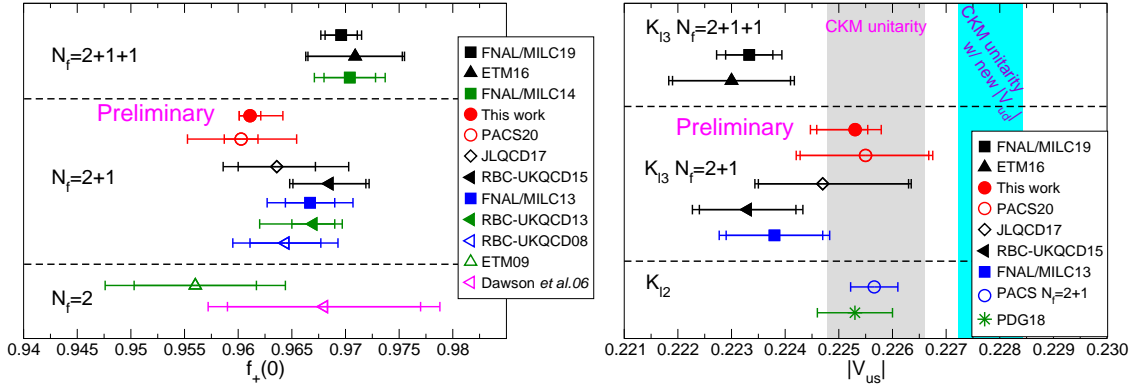

 \centering
 \includegraphics*[scale=0.40]{figs/f00.eps}
 \includegraphics*[scale=0.40]{figs/vus.eps}
 \caption{
Left: Our preliminary result of $f_+(0)$ in the continuum limit
is plotted by the closed circle,
together with the result in our previous calculation~\cite{PACS:2019hxd} 
and those obtained by other groups~\cite{Dawson:2006qc,Boyle:2007qe,Lubicz:2009ht,Bazavov:2012cd,Boyle:2013gsa,Bazavov:2013maa,Boyle:2015hfa,Carrasco:2016kpy,Aoki:2017spo,Bazavov:2018kjg}.
The closed and open symbols denote results in the continuum limit and
at one lattice spacing, respectively.
The inner and outer errors express the statistical and total errors.
The total error is evaluated by adding the statistical and systematic
errors in quadrature.
Right: Result of $|V_{us}|$. The symbols are the same in the left panel
except for the values from the $K_{l2}$ decay process.
Those are determined with $F_K/F_\pi$ in PDG18~\cite{Tanabashi:2018oca}
(star) and from the PACS10 configuration at 
$a = 0.085$ fm~\cite{Ishikawa:2018jee}.
The gray and light blue bands express the values estimated through
the unitarity condition of the CKM matrix using the traditional value of 
$|V_{ud}|$~\cite{Marciano:2005ec,Tanabashi:2018oca} 
and the updated one~\cite{Seng:2018yzq}, respectively.
The inner and outer errors express the lattice and total errors.
The total error is evaluated by adding the lattice and experimental
errors in quadrature.
  \label{fig:f00_vus}
 }
\end{figure}

The result of $f_+(0)$ in the continuum limit 
is plotted in the left panel of Fig.~\ref{fig:f00_vus}.
Its central value and the statistical error are determined from the fit A.
The systematic error is estimated from the difference of
the results for the fit A and B.
The preliminary result in this calculation is consistent with
that in our previous calculation~\cite{PACS:2019hxd}, which is obtained
from only the local current data at $a = 0.085$ fm
plotted by the open circle in the figure.
In this calculation the statistical and systematic errors are improved
compared with the previous ones.
Our result is roughly consistent with other lattice results
in $N_f = 2$ and 2+1 QCD~\cite{Dawson:2006qc,Lubicz:2009ht,Boyle:2007qe,Bazavov:2012cd,Boyle:2013gsa,Boyle:2015hfa,Aoki:2017spo},
while it is slightly smaller than those in $N_f = 2+1+1$ QCD~\cite{Bazavov:2013maa,Carrasco:2016kpy,Bazavov:2018kjg}.
At present the reasons are not clear, and we would need detail investigations
for the difference in future.

Using the experimental value 
$|V_{us}|f_+(0) = 0.21654(41)$~\cite{Moulson:2017ive},
we determine the value of $|V_{us}|$ with our preliminary result,
which is plotted in the right panel of Fig.~\ref{fig:f00_vus}.
The total error contains the statistical and systematic errors in 
our calculation and also the error from the experiment.
For comparison, results from other groups are also plotted in the figure.
Our result agrees with the value of $|V_{us}|$
determined from the kaon leptonic ($K_{l2}$) decay process through
$|V_{us}|/|V_{ud}| \times F_K/F_\pi = 0.27599(38)$~\cite{Tanabashi:2018oca}
using the value of $F_K/F_\pi$ in PDG18~\cite{Tanabashi:2018oca}.
We also observe a consistency of our result with
the one using the result of $F_K/F_\pi$ calculated 
with the PACS10 configuration at $a = 0.085$ fm~\cite{Ishikawa:2018jee}.

Our preliminary result is also in good agreement with
$|V_{us}|$ estimated from the unitarity condition of 
the first row of the CKM matrix using the value of $|V_{ud}|$ in 
Refs.~\cite{Marciano:2005ec,Tanabashi:2018oca},
$|V_{us}| = \sqrt{ 1 - |V_{ud}|^2 }$
where $|V_{ub}|$ is neglected due to $|V_{ub}| \ll |V_{ud}|$.
On the other hand,
when we use the updated value of $|V_{ud}|$~\cite{Seng:2018yzq},
a clear difference is seen between our preliminary result
and the estimated value from the unitarity expressed 
by the light blue band in the right panel of Fig.~\ref{fig:f00_vus}.
This might be a signal of new physics beyond the standard model,
while it needs to be investigated with a more precise result.

\section{Summary}

We have calculated the $K_{l3}$ form factors with the PACS10 configurations
at $a = 0.064$ and 0.085 fm.
In addition to the local current calculation as in our previous work,
the conserved current is employed in the calculation 
in order to carry out the continuum extrapolation of the form factors.
For the continuum extrapolation, simultaneous fits are performed with
all the data for the form factors using the fit forms based on 
the NLO SU(3) ChPT formulas with terms for finite lattice spacing effects.

Our preliminary result of $f_+(0)$ in the continuum limit agrees with
our previous work at $a = 0.085$ fm.
Our result also reasonably agree with
other lattice results except for those in $N_f = 2+1+1$ QCD, which are
little larger than our value.
We would need to investigate reasons of the difference in future.
Using our preliminary result of $f_+(0)$, the value of $|V_{us}|$
is determined.
As in our previous calculation, it is consistent with the values
obtained through the $K_{l2}$ decay, although there is a
deviation from the estimated value with the CKM unitarity using
the updated $|V_{ud}|$.

In this study we have estimated a systematic error coming from
only the continuum extrapolation, so that other systematic errors need
to be estimated.
Furthermore, since the continuum extrapolation is performed with 
the data in only two different lattice spacings, 
at least another lattice spacing data is
necessary for a more reliable result in the continuum limit.
Towards this direction, we are generating the third PACS10 configuration
at a finer lattice spacing.

\section*{Acknowledgments}
Numerical calculations in this work were performed on Oakforest-PACS
in Joint Center for Advanced High Performance Computing (JCAHPC)
under Multidisciplinary Cooperative Research Program of Center for Computational Sciences, University of Tsukuba.
This research also used computational resources of the HPCI system provided by Information Technology Center of the University of Tokyo and RIKEN CCS through the HPCI System Research Project (Project ID: hp170022, hp180051, hp190081, hp200062, hp200167, hp210112).
The calculation employed OpenQCD system\footnote{http://luscher.web.cern.ch/luscher/openQCD/}.
This work was supported in part by Grants-in-Aid 
for Scientific Research from the Ministry of Education, Culture, Sports, 
Science and Technology (Nos. 16H06002, 18K03638, 19H01892).

\bibliography{reference}

\end{document}